\documentclass[a4paper,11pt,twocolumn,accepted=2020-02-03]{quantumarticle}

\pdfoutput=1
\usepackage[utf8]{inputenc}
\usepackage[english]{babel}
\usepackage[T1]{fontenc}
\usepackage{hyperref}

\usepackage{tikz}
\usepackage{lipsum}

\usepackage{amssymb,amsmath,amsfonts}
\usepackage[normalem]{ulem}
\usepackage{bm,graphicx}
\usepackage{dsfont}
\usepackage{mathrsfs}
\usepackage[none]{hyphenat}

\usepackage[mathscr]{eucal}
\usepackage{color}
\usepackage{marvosym}  

\usepackage[caption=false]{subfig}
\DeclareInputText{175}{\.Z}     

\newcommand{\id}{\mathds{1}}

\DeclareMathOperator{\Tr}{Tr}

\newcommand{\ket}[1]{\left|{#1}\right\rangle}
\newcommand{\bra}[1]{\left\langle{#1}\right|}

\newcommand{\ketbrad}[1]{\left|{#1}\rangle\!\langle{#1}\right|}

\newcommand{\rme}{\ensuremath{\mathrm{e}}}
\newcommand{\rmi}{\ensuremath{\mathrm{i}}}

\newcommand{\gd}{\ensuremath{\mathsf{g}}}

\begin{document}

\title{Maximum $N$-body correlations do not in general imply 
       genuine multipartite entanglement}

\author{Christopher Eltschka} 
\affiliation{Institut f\"ur Theoretische Physik, Universit\"at Regensburg, D-93040 Regensburg, Germany
            }

\author{Jens Siewert} 
\affiliation{Departamento de Qu\'{i}mica F\'{i}sica, Universidad del Pa\'{i}s Vasco UPV/EHU, E-48080 Bilbao, Spain}
\affiliation{IKERBASQUE Basque Foundation for Science, E-48013 Bilbao, Spain
}

\maketitle

\begin{abstract}
     The existence of correlations between the parts of a quantum
     system on the one hand, and entanglement between them on the other,
     are different properties.
     Yet, one intuitively would identify strong $N$-party correlations with
     $N$-party entanglement in an $N$-partite quantum state. 
     If the local systems are qubits, this intuition
     is confirmed: The state with the strongest $N$-party correlations
     is the Greenberger-Horne-Zeilinger (GHZ) state, which does have genuine
     multipartite entanglement. 
     However, for high-dimensional local systems the state with strongest
     $N$-party correlations may be 
     a tensor product of Bell states, that is, partially separable.
     We show this by introducing several novel tools for handling the
     Bloch representation.
\end{abstract}

\section{Introduction}%
%
The expansion of the density  operator 
in terms of a matrix basis
is called the Bloch representation~\cite{Fano1954,Fano1957,Mahler2004}. 
Technically, this representation
is rather demanding: A pure state of $N$ parties each of local
dimension $d$ is characterized by $2d^N-2$  real coefficients, whereas
the same state written in the Bloch representation requires $d^{2N}-1$
parameters, just as any mixed state. 
On the other hand, this 
representation appears to be perfectly adapted to studying the 
correlation properties of a
quantum system, because  the above-mentioned expansion corresponds to
a decomposition of the state into all possible correlation contributions
(hence the terms are also called correlation tensors).
Therefore, from a better understanding of the technical characteristics of
this expansion one may expect significant insight into the physics
of correlated quantum systems.

The systematic investigation of the properties of the Bloch representation
for finite-dimensional multi-party quantum systems is a relatively 
recent subject~\cite{Kloeckl2015,ES2015,Tran2016,Appel2017,Huber2017,ES2018,Wyderka2018,Huber2018,Eltschka2018,Cox2018,Wyderka2019,ES2019}, although many important results were found earlier,
mostly relating specific features of the Bloch picture to the 
entanglement properties of the state 
(e.g.,~\cite{Mahler1995,Mahler1996,Zukowski2002,Jaeger2003,Aschauer2004,Scott2004,deVicente2007,deVicente2008,Badziag2008,HuberdeVicente2011,Laskowski2011}). Currently much activity is devoted to working out the technical details
and properties for an easier use of the Bloch representation to solving
physics problems. An essential part of this is to figure out smaller sets
of parameters that carry sufficient amounts of information to facilitate
the characterization of relevant physical properties for a state given in the 
Bloch representation. In this contribution, we define such a set of parameters,
which we call the ``sector distribution'' 
and discuss some of its key features. Moreover, we illustrate how the 
properties of this distribution are reflected in the correlation properties
of the states.

To be more specific, let us preliminarily introduce
the Bloch representation; the precise definition will be given 
below. If we enumerate the parties of an $N$-party system (of equal
local dimension $d$) by $\{1,2 \ldots N\}$ and $A$ is a 
subset of parties, then
\begin{align}
  \rho\ =\ \frac{1}{d^N}\sum_{A} G_A\otimes \id_{\bar{A}}\ \ .
\label{eq:firstBloch}
\end{align}
Here, $G_A$ is a Hermitian operator that acts nontrivially on the
parties belonging to the subset $A$, and $\id_{\bar{A}}$ is the identity
operator for the complementary set. 
Consider now the sum of all those
terms in Eq.~\eqref{eq:firstBloch} that act on the same number $k$
of parties, 
\begin{align}
   \mathscr{S}_k\ \equiv\ \frac{1}{d^N} 
                          \sum_{|A|=k} G_{A}\otimes\id_{\bar{A}}\ \ .
\label{eq:sect}
\end{align}
We call $\mathscr{S}_k$ the ``$k$-sector'' of $\rho$ and the 
(squared) Hilbert-Schmidt length of $\mathscr{S}_k$ the ``$k$-sector length''
$S_k$~\cite{footnote1}
\begin{align}
    S_k\ \equiv\ d^N
                 \Tr\left(\mathscr{S}_k^{\dagger}\mathscr{S}_k\right) \ \ .
\end{align}
The $k$-sector length $S_k$ is a natural 
quantifier for the $k$-party correlations in a state~\cite{Girolami2017}.
Clearly, for an $N$-partite state there are $N$ sector lengths
($\mathscr{S}_0=1$ for all normalized states). 
Sector lengths were discussed earlier~\cite{Mahler1995,Mahler1996,Aschauer2004,Kloeckl2015,ES2015,Tran2016,Huber2017,Appel2017,ES2018,Wyderka2018,Wyderka2019,ES2019}.
In particular the $N$-sector was intuitively linked with the $N$-party
quantum correlations. Therefore it came as a surprise that there exist mixed
states that are $N$-party entangled but do not possess $N$-party 
correlations~\cite{Kaszlikowski2008,Schwemmer2015,Tran2017,Klobus2019}. 
Later it was realized~\cite{Kloeckl2015} 
that, in order to witness genuine multipartite entanglement, 
it may be necessary to consider a collection of the highest sector 
lengths $S_N, S_{N-1},\ldots$ rather than just $S_N$.
In the present work we systematically study the set of
all sector lengths $\{S_k\}$.
As we will demonstrate
the distribution $\{S_1,S_2\ldots S_N\}$ represents a reduced
set of parameters in the spirit described above
(linear in the system size instead of exponential)
that carries substantial information regarding some of the correlation
properties of the state. Often it has little meaning to 
study the sector lengths $S_k$ individually; rather, there exists a
variety of strict
relations between them that determine the entire distribution.

We introduce several novel technical concepts, most importantly
the $N$-sector projector. We use this toolbox to prove the long-standing
conjecture that for any number $N$ of qubits the GHZ state maximizes the 
$N$-sector length. Subsequently we analyze the sector distribution for few-party
systems of higher local dimension $d>2$. Here we prove that for higher
local dimension the state with maximum $N$-sector length may be 
partially separable.
Moreover, we provide a comprehensive discussion for the behavior of 
the $N$-sector
with increasing number of parties as well as growing local dimension.
%
%

\section{Definitions and preliminaries}
%
For the Bloch representation of an $N$-partite state with
all local dimensions equal to $d\geqq 2$ the operators $G_A$ in 
Eq.~\eqref{eq:firstBloch} are expanded in local operators, i.e.,
a basis of traceless matrices $\{\gd_j\}$, 
$1 \leqq j \leqq d^2-1$, $\gd_0\equiv \id$,
with normalization
$\Tr\left(\gd_j^{\dagger}\gd_k\right)=d\delta_{jk}$, 
\begin{align}
   \rho = \frac{1}{d^N}\sum_A\sum_{ \scriptsize
                                            j_l:\ l\in A
                                          }
            \!\!
            r_{j_1\cdots j_N}\
            \gd_{j_1}\otimes\cdots\otimes\gd_{j_N}\otimes \id_{\bar{A}}
\  ,
\label{eq:Bloch}
\end{align}
and
\begin{align}
            r_{j_1\cdots j_N}\ =\ \Tr\left(
              \left[
              \gd^{\dagger}_{j_1}\otimes\cdots\otimes\gd_{j_N}^{\dagger}\otimes
                       \id_{\bar{A}}
              \right]\rho\right)\ \ .
\end{align}
Here, all indices $j_m$, $m\in \bar{A}$ 
are set to 0.
With this, the $k$-sector length simply becomes
\begin{align}
     S_k\ =\ \sum_{ j_l:\ l\in A, |A|=k } |r_{j_1\cdots j_N}|^2
\ \ ,
\end{align}
where $|A|$ is the number of elements in $A$.
All the sector lengths are local unitary invariants of the state.
As the actual sector lengths $\sqrt{S_k}$ equal the Hilbert-Schmidt
norms of $\mathscr{S}_k$, they obey the triangle inequality. 
Because of this, the $N$-sector length of a mixture of pure states
can never exceed the largest $N$-sector length of any of the pure states.
Therefore,
throughout this article we focus on pure states $\Pi$, that is, for 
the purity we have $\Tr\Pi^2=\Tr\Pi=1$ and, hence, for the sum of
all sector lengths, $\sum_{k=0}^N S_{k}=d^N$.

Consider the simplest case of all, that is, product states 
$\ket{\mathsf{prod}_j^N}=\ket{j}^{\otimes N}$; here $\ket{j}$ denotes
a state of the computational basis, $j=0,1, \ldots, (d-1)$. It is 
easy to see that \mbox{$S_k(\mathsf{prod}_j^N)={N\choose k}(d-1)^k$}. 
Remarkably, it was
shown by Tran {\em et al.}~\cite{Tran2016} that among the pure
states only product states have the minimum $N$-sector length 
\[
\min S_N=(d-1)^N\ \ , 
\]
that is, for $d>2$ the 
$N$-sector is always on the order of $d^N$.
It turned out
that the opposite question regarding the states with {\em maximum} $N$-sector
is considerably more complex. A relevant state for this discussion is
the GHZ state for $N$ parties of local dimension $d$
\begin{align}
 \ket{\text{GHZ}_d^N}\ =\ \frac{1}{\sqrt{d}}\sum_{j=0}^{d-1} \ket{j}^{\otimes N}
\label{eq:ghz}
\ \ .
\end{align}
For $N$-qudit GHZ states we find the sector distribution (see Appendix)
%
%
\begin{align}
    S_k(\text{GHZ}_d^N)\ =\ & (d-1)d^{N-1}\delta_{kN}\ +
\nonumber\\
                            & + \ {N\choose k}\frac{(d-1)^k+(-1)^k(d-1)}{d}
\  .
\label{eq:ghz-sectors}
\end{align}

Tran {\em et al.}~\cite{Tran2016} showed 
that for odd party number $N$ of qubits  the GHZ state has 
maximum $N$-sector length. They conjectured that this statement
holds also for even $N$. For $d>2$ it is not clear which state
has maximum $N$-party correlations. 
In the following we will prove the conjecture for even-$N$
qubit GHZ states.
%

\section{The $N$-sector projector}
%
For the proof we need some mathematical tools based on universal
state inversion~\cite{Rungta2001,Hall2005,Lewenstein2016,ES2018,Eltschka2018}.
First we define the projection (super-)operator~\cite{footnote2}
onto the last $(k=N)$ sector (or $N$-sector projector for short 
whenever there are no ambiguities)
\begin{align}
    \mathcal{P}(\rho)\ =\ 
                                  \prod_{j=1}^N \left[ \mathsf{id}-\frac{1}{d}
                                  \Tr_j(\cdot)\otimes\id_j\right] \rho
\ \ .
\label{eq:P}
\end{align}
It is easy to check by writing $\rho$ in the Bloch representation that
$\mathcal{P}$ indeed realizes a projection onto the $N$-sector, $\mathscr{S}_N$.
The map~\eqref{eq:P} belongs to the class of generalized
universal state inversions discussed in 
Refs.~\cite{Lewenstein2016,Eltschka2018} that have the form 
$\mathcal{I}^{\{\alpha_j,\beta_j\}}=\prod_{j=1}^N 
 \big[\alpha_j\Tr_j(\cdot)\otimes\id_j-\beta_j\mathsf{id}\big]$, 
where $\mathsf{id}$
denotes the identity map and $\alpha_j$, $\beta_j$ are real numbers. 
With definition~\eqref{eq:P} we get immediately
\begin{align}
   S_N(\rho)\ =\ & d^N\Tr\left[\rho\ \mathcal{P}(\rho)\right]
\nonumber\\
        =\ &d^N\sum_{A} \left(-\frac{1}{d}\right)^{|A|} 
                            \Tr\left(\rho_{\bar{A}}^2\right)\ ,
\label{eq:NSecPurity}
\end{align}
where $A$, as before, runs through all subsets of $\{1\ldots N\}$
and $\rho_{\bar{A}}=\Tr_{A}\rho$ is the reduced state on the subset of 
parties $\bar{A}$. The equality $\Tr\left[\left(\Tr_A \Pi\right)^2\right]=
                          \Tr\left[\left(\Tr_{\bar{A}} \Pi\right)^2\right]$
for pure states $\Pi$ motivates the definition of another operator
\begin{align}
    \mathcal{Q}(\rho)\ =\ \prod_{j=1}^N \left[ \Tr_j(\cdot)\otimes\id_j
                                           -\frac{1}{d}\mathsf{id}
                                             \right] \rho
\ \ ,
\label{eq:Q}
\end{align}
so that 
\begin{align}
   S_N\ =\ & d^N\Tr\left[\Pi\ \mathcal{P}(\Pi)\right]
\nonumber\\
   =\ & d^N\Tr\left[\Pi\ \mathcal{Q}(\Pi)\right]\ \geqq\  0\  .
\label{eq:P=Q}
\end{align}
Because of the projector property of $\mathcal{P}$ and the
Cauchy-Schwarz inequality, 
$\|M\|_{\mathcal{P}}\equiv \sqrt{\Tr\left[M^{\dagger} \mathcal{P}(M)\right]}$
defines a seminorm for operators $M$ (while the analogous statement
does not hold for $\mathcal{Q}$).
By considering the action of $\mathcal{Q}(\Pi)$ in the Bloch
representation relation~\eqref{eq:P=Q} gives rise to the astounding equality
%
\begin{align}
  d^N  S_N(\Pi)\ =\  \sum_{k=0}^N  (-1)^k(d^2-1)^{N-k} S_k(\Pi)
\ \ ,
\label{eq:PQrel}
\end{align}
which links the last sector length $S_N$ with all others. 

Finally we rewrite the well-known purity condition for reductions
of pure states $\Tr\left[\left(\Tr_A\ketbrad{\psi}\right)^2\right]=
\Tr\left[\left(\Tr_{\bar{A}}\ketbrad{\psi}\right)^2\right]$ 
in terms of sector lengths.
This is achieved by symmetrizing the purity conditions for fixed $|A|$ and
accomplishing the combinatorial accounting. We find for the $k$-purity relation
($k=0,1,\ldots, \lfloor{\frac{N-1}{2}\rfloor}$ with the 
floor \mbox{function 
$\lfloor{\cdot}\rfloor$)~\cite{Felix}}
\begin{align}
 d^{N-2k}   \sum_{m=0}^k {N-m\choose k-m} S_m\ =\
 \sum_{n=0}^{N-k} {N-n\choose k} S_n\ \ .
\label{eq:k-purity}         
\end{align}
For $k=0$ this gives the well-known condition $d^N=\sum_0^N S_n$.
We explicitly write the relations for $k=1$ and $k=2$ as they will
turn out useful later:
\begin{widetext}
\begin{subequations}
\begin{align}
  d^{N-2}\left[ N+S_1 \right]\ = \ & N + (N-1)S_1 + \ldots + 2 S_{N-2} +
                                     S_{N-1}\ \ ,
\label{eq:1-purity}
\\
  d^{N-4}  \left[{N\choose 2} + (N-1) S_1 +  S_2
         \right]            \ = \ &
                 {N\choose 2}+{N-1\choose 2} S_1 +\ldots+{3\choose 2}S_{N-3}+
                    S_{N-2}
\ \ .
\label{eq:2-purity}
\end{align}
\label{eq:purity}
\end{subequations}
\end{widetext}
Interestingly, Eqs.~\eqref{eq:k-purity} elucidate the role of the
$N$-sector for pure states: All sector lengths have to be adjusted so as to
obey the $k$-purity relations ($k=1 \ldots \lfloor \frac{N-1}{2} 
\rfloor$)
between the reduced states of non-empty complementary partitions; 
note that the last sector is {\em excluded}
from establishing this balance. The last sector serves 
to fill up the total length $d^N$ of the Bloch vector (the ``0-purity''
relation).

\section{The $N$-qubit GHZ state maximizes the $N$-sector}
%
For odd $N$ we know 
$\max S_N=S_N(\text{GHZ})=2^{N-1}$, 
cf.~Ref.~\cite{Tran2016}. We recall this proof in the Appendix. 
We are prepared now to show for even $N$ qubits that 
\[
\max S_N=S_N(\text{GHZ})=2^{N-1}+1\ \ .
\]
As here the GHZ state has only even-numbered sectors, Eq.~\eqref{eq:PQrel} would
imply for a hypothetical state $\Pi_x=\ket{x}\!\bra{x}$ with 
larger $N$-sector than
GHZ that 
$S_{2m}>{N\choose 2m}$ for some $m<N/2$. 
In order to obtain information
regarding the distribution of the even-numbered sectors we
consider an $R$ matrix (analogous to Refs.~\cite{CKW2000,ES2015,ES2018})
of $\rho_{[1]}\equiv\Tr_{\{1\}}\Pi_x$ after tracing the first party,
%
\begin{align}
    R_{[1]} \equiv \rho_{[1]} \mathcal{I}_-(\rho_{[1]})
            = \rho_{[1]} \sum_A\! (-1)^{|A|}\Tr_{\bar{A}} \rho_{[1]}\otimes
                                        \id_{\bar{A}} \ ,
\label{eq:Rmat}
\end{align}
where 
$\mathcal{I}_-(\sigma)\equiv \prod_{j=1}^M 
    \left[\Tr_j(\cdot)\otimes \id_j-\mathsf{id}\right]\sigma$ 
denotes the standard universal state inversion for an $M$-partite
state $\sigma$ (cf.~Refs.~\cite{Hall2005,ES2018,Huber2018}).
If we symmetrize over the traced party we can establish
a relation between $\sum_j\Tr R_{[j]}$  and the sector lengths 
$S_k(\Pi_x)$ of $\Pi_x$,
%
\begin{align}
 0\ \leqq\   d^{N-1} \sum_{j=1}^N\Tr R_{[j]}\ =\ \sum_{k=0}^{N-1} (-1)^k(N-k)S_k
\ \ .
\label{eq:TrRsymm}
\end{align}
%
The reasoning is exactly the same as the one to obtain the 1-purity,
Eq.~\eqref{eq:1-purity}. By adding the latter equation and 
Eq.~\eqref{eq:TrRsymm} (and dividing by 2)
we obtain a relation for the even-numbered sectors, 
\begin{widetext}
\begin{align}
  \frac{d^{N-2}}{2}\left[N\ +\ S_1\ +\ \sum_jd\Tr R_{[j]}
         \right]\ =\ N + (N-2) S_2 + (N-4) S_4 +\ldots
                   + 4 S_{N-4} + 2 S_{N-2}\ \ .
\label{eq:evenlin}
\end{align}
\end{widetext}
We observe that on the right-hand side (r.h.s.) the prefactors increase with
decreasing index, this is analogous to Eq.~\eqref{eq:PQrel},
only that here the prefactors increase linearly.
Also here the even-sector distribution of $\Pi_x$
would exceed the result of the GHZ state.
Relation~\eqref{eq:evenlin} gives us the possibility to directly
check the achievable maximum of the r.h.s.\ for pure states by maximizing the
terms on the left-hand side. For qubits this is straightforward and shows
that the maximum is achieved for the GHZ state (we present this calculation
in the Appendix). Hence, there
is no state $\Pi_x$ with larger $N$-sector.\hfill $\square$
%
%

%
\section{Few parties of higher local dimension}
%
The obvious guess from the results so far is that $\ket{\text{GHZ}_d^{N}}$ 
maximizes the $N$-sector length also for $d>2$. 
It will turn out that this can only partially be true.
To this end, let us investigate states with up to six parties. 
The following results are obtained by using 
Eqs.~\eqref{eq:PQrel},~\eqref{eq:k-purity} 
for $k=0,1,2$, and increasingly tedious algebra.
\\
$N=2:$ We have $d^2=1+S_1+S_2$, so that 
\[
\max S_2= d^2-1 
\]
for $S_1=0$, that is, 
the Bell state $\ket{\Phi_d^+}\equiv\ket{\text{GHZ}_d^2}$
maximizes the 2-sector.\\
$N=3$: Here, 
\[   S_3=(d-1)^2(d+2)-(d-1)S_1\ \ ,
\]
so that $S_1=0$ leads to $\max S_3=(d+2)(d-1)^2$, which again is realized
by the GHZ state.\\
$N=4$: In this case, there remains more than one parameter undetermined
\[
  S_4=(d^2-1)^2-\frac{1}{2}\left[ (d^2-1)S_1+S_3\right]
\ \ ,
\]
but since $S_1, S_3\geqq 0$, the 4-sector gets maximized for 
$S_1=S_3=0$, so that $\max S_4=(d^2-1)^2$. That is, for 
four-party states the $N$-sector is {\em not} maximized by
the GHZ state, but by a tensor product of Bell states, i.e., a
biseparable state.
Curiously, the case $d=2$ is right on
the edge, because the tensor product of a pair of two-qubit Bell states and
the four-qubit GHZ state have the same 4-sector length, $S_4=9$.\\
$N=5$: Here we find
\begin{align}
   (d-3)S_5 & \ =\  (d-1)^3(d+2)(d^2-2d-4) -
\nonumber\\
                & - (d-1)^2(d^2-d-3)S_1 + (d-1)S_3\ \ .
\label{eq:N=5}
\end{align}
This suggests again $S_1=0$ 
for the $N$-sector maximum, however, now the
sign of $S_3$ is reversed. We note that the maximum $k$-sector length 
of an $N$-party system is on the order of ${N\choose k}d^k$, so that 
the 3-sector length $S_3\sim O(d^3)$, and hence for
large local dimension $d\gg 1$ 
\[   \frac{\max S_5}{(d-1)^3(d+1)(d+2)}\ \longrightarrow\ 1\ 
                                              +\ O(d^{-2})\ \ ,
\]
%
which indicates that the maximum $N$-sector is approximated with
better than first-order accuracy for growing $d$ by the polynomial 
in the denominator.
The latter corresponds to the tensor product of a Bell state and a
three-party GHZ state, $\ket{\Phi_d^+}\otimes\ket{\text{GHZ}_d^3}$. 
Consequently, for large $d$ also here the state with maximum $N$-body
correlations may be biseparable. The case $d=3$ is special: $S_3=20$
gives the largest 3-sector. The five-qutrit GHZ state is compatible with
this [cf.~Eq.~\eqref{eq:ghz-sectors}] and has larger 5-sector than 
$\ket{\Phi_3^+}\otimes\ket{\text{GHZ}_3^3}$ (172 vs.\ 160). However,
in principle, there might be a state with $S_1=0$, $S_3=20$ 
and even larger 5-sector.\\
$N=6$: This case has similar features as $N=5$. The 6-sector obeys
\begin{align}
 2(d^2-4)S_6\ =\ & 2(d-2)(d^2-1)^3(d+2)\ -
\nonumber\\
               & -(d^2-1)^2(d^2-3)S_1\  +
\nonumber\\
               & + (d^2-1)S_3-(d^2-3)S_5
\ \ .
\label{eq:N=6}
\end{align}
Again we see that for increasing $d\gg 1$ the 6-sector 
$(\max S_6)/(d^2-1)^3\longrightarrow 1+O(d^{-3})$ 
because of the scaling of the sector lengths
with $d$; the corresponding state is $\ket{\Phi_d^+}^{\otimes 3}$. 
Note that already
for $d=3$ the 6-sector of $\ket{\Phi_3^+}^{\otimes 3}$ beats the length
of the six-qutrit GHZ state (512 vs.\ 508). 

We summarize the results of this section in Tables~1--3.
%
\begin{table}[h]
\caption{Maximum of the $N$-sector for 2, 3, 4 parties.}
\begin{tabular}{ccc}
\hline
\hline
$N$  & max $N$-sector $S_N$  &  state maximizing $S_N$  \\[.9mm] \hline
 2   & $d^2-1$               & $\ket{\Phi_d^+}$   \\[.9mm]
 3   & $(d-1)^2(d+2)$        & $\ket{\mathrm{GHZ}^3_d}$   \\[1.0mm]
 4   & $(d^2-1)^2$           & $\ket{\Phi_d^+}^{\otimes 2}$   \\[.7mm]
\hline
\hline
\end{tabular}
\end{table}
%
\begin{table}[h]
\caption{Comparison of $N$-sectors for $N=5$.}
\begin{tabular}{cccc}
\hline
\hline
$S_5$  &  $\ket{\mathrm{GHZ}^5_d}$     &  &
                   $\ket{\mathrm{GHZ}^3_d}\otimes\ket{\Phi_d^+}$      \\[.4mm]
$d$    &  $(d-1)^2 \times $       & &      $(d-1)^3\times $           \\[.4mm]
           &   $(d^3+2d^2-2d+4)$      &  & $ (d+1)(d+2)$
\\[.4mm] \hline
 3   & 172   &  $\mathbf{>} $   & 160
\\[.4mm]
 4   & 828      &  $\mathbf{>} $  & 810
\\[.4mm]
 5   &  2704     &  $>$   &    2688
\\[.4mm]
 6   &  7000     &  $=$   &    7000
\\[.4mm]
 7   &  15516     & $<$   &   15552
\\[.4mm]
\hline
\hline
\end{tabular}
\end{table}
%
\begin{table}[h]
\caption{Comparison of $N$-sectors for $N=6$.}
\begin{tabular}{cccc}
\hline
\hline
$S_6$  &  $\ket{\mathrm{GHZ}^6_d}$     &  &
                   $\ket{\Phi_d^+}^{\otimes 3}$      \\[.4mm]
$d$      &  $\frac{d-1}{d}\left[d^6+(d-1)^5+1\right]$      
                  &  & $(d^2-1)^3 $
\\[.4mm] \hline
 2   & 33   &  $\mathbf{>} $   & 27
\\[.4mm]
 3   & 508      &  $\mathbf{<} $  & 512
\\[.4mm]
 4   &  3255     &  $<$   &    3375
\\[.4mm]
\hline
\hline
\end{tabular}
\end{table}
%
%
\begin{figure}[th]
  \centering
  \includegraphics[width=.99\linewidth]{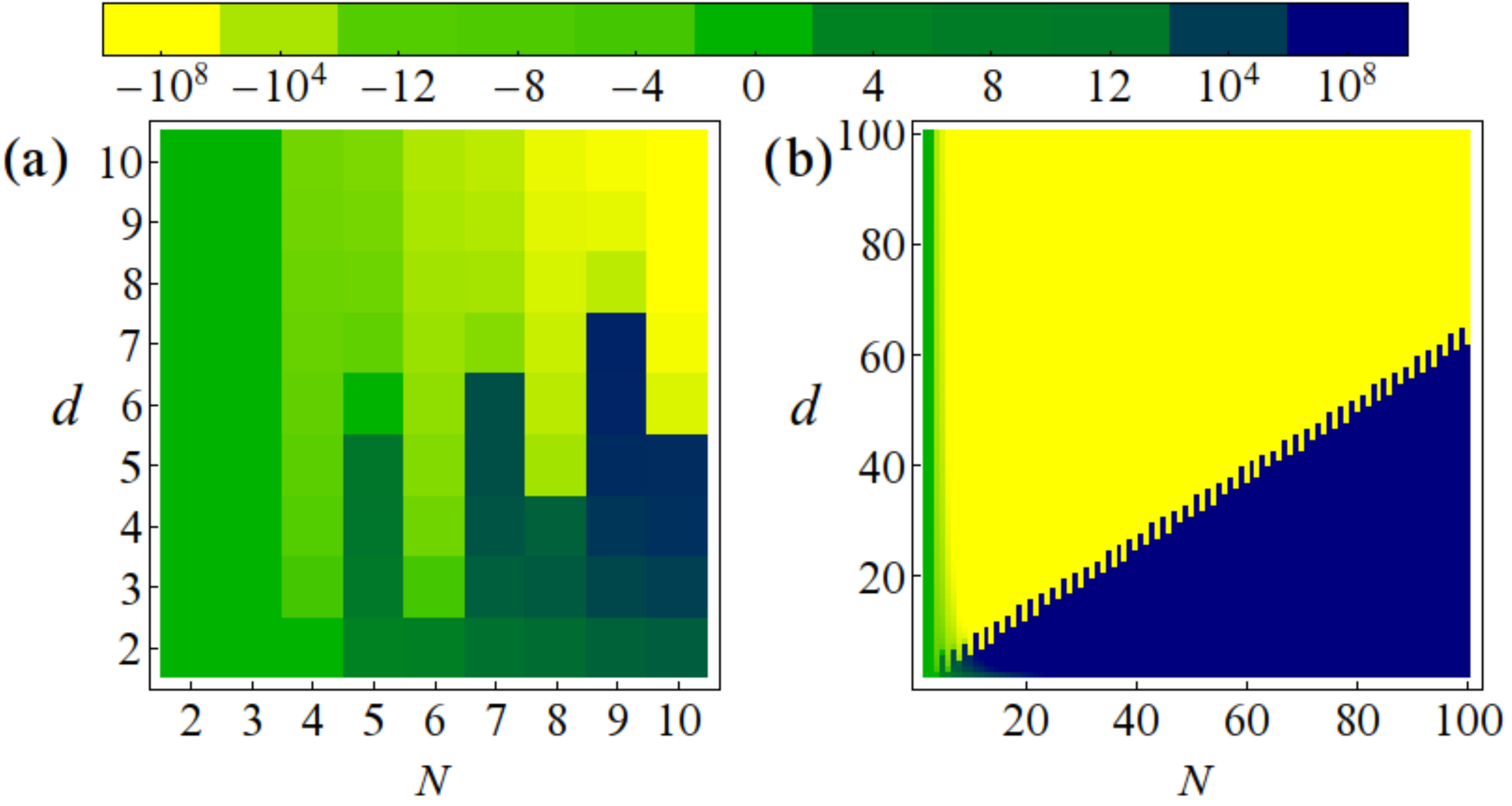}
  \caption{$N$-sector length
           difference $S_N(\text{GHZ}_d^N)-S_N(\mathsf{Bell}_d^N)$.
           The border between GHZ-dominated and {\sf Bell}-dominated
           is given by a straight line $d\simeq 0.6275\cdot N$ (see text);
           however, note the pronounced even-odd effect.
           For $N=2$ and $N=3$, GHZ and {\sf Bell} are the same state, 
           therefore these cases have to be counted as `undecided'.
           (a) Small scale $N,d\leqq 10$; (b) larger scale $N,d\leqq 100$.
           Note that the color scale is logarithmic.
    }
  \label{fig:fig1}
\end{figure}

%

%
%
%
\section{Maximum $N$-sector for large $d$ and large $N$}
%
We can investigate the dominance of $N$-sectors numerically.
On increasing $d$, the partially separable states---that is,
a tensor product of Bell states (even $N$)
or Bell states and a three-party GHZ state (odd $N$)---appear to dominate
(we will call these states ``$\mathsf{Bell}$''). In the opposite case,
the GHZ state has larger $N$-sector.
This behavior is shown in Fig.~1,
where the difference of $N$-sectors 
$S_N(\text{GHZ}_d^N)-S_N(\mathsf{Bell}_d^N)$
is displayed in a $(d,N)$ plane. We note 
 the formal analogy of our problem 
of finding the maximum $N$-sector with that of deciding the existence
of absolutely maximimally entangled (AME)
states~\cite{Scott2004,Goyeneche2014,Alsina2015,Huber2017,Huber2018}
that is suggested by the structure of Fig.~1. 
The analogy arises because also the $N$-sector problem seems to 
have two solutions
whose validity regions in the $(d,N)$ plane 
are connected and separated by a single line.
The region of GHZ dominance
corresponds to `AME state does not exist', while that of {\sf Bell}
dominance relates to `AME does exist'. The line separating the 
two corresponds to the Scott bound~\cite{Scott2004,Huber2018}.
An accurate analytical approximation for this line is found by equating
$S_N(\text{GHZ}_d^N)$ in Eq.~\eqref{eq:ghz-sectors} with 
$S_N(\mathsf{Bell}_d^N)=(d^2-1)^{N/2}$ (for even $N$). 
Assuming $d=\gamma^{-1} N$,
this leads to an equation that determines the $\gamma$ parameter,
$\rme^{-\gamma}=1-\frac{\gamma}{2}$, from which
\begin{equation}
d\simeq 0.6275\cdot N\ \ .
\end{equation}
Close to this line there may be exceptions from 
the rule, just as in the case of AME states.

In the following we provide arguments why GHZ and {\sf Bell}
are, if not the dominating, at least close to the states with
dominating $N$-sector in the limits of large $N$ and $d$.
Consider first fixed {\em even} $N$ and $d\gg 2, N$. 
Our reasoning is based on the purity relations Eqs.~\eqref{eq:purity}
and on the consideration that to leading order 
the maximum $k$-sector is given by $S_k\sim {N\choose k}d^k$.
From the 0-purity relation
$S_N=d^N-S_{N-1}-\ldots -1$ it follows that the dominating terms
$S_{N-1}+S_{N-2}$ need to be as small as possible in order to obtain 
$\max S_N$. 
We observe 
that Eq.~\eqref{eq:1-purity} dictates that 
$S_{N-2}$ and $S_{N-1}$  cannot both vanish, and their sum needs to 
be at least of order $N d^{N-2}$.
As $S_1 > 0$ would only increase the r.h.s., $S_1=0$ 
is the sensible choice.
Moreover,
we see that $S_{N-2}+S_{N-1}\approx 
                        \frac{1}{2}\left(d^{N-2}N+S_{N-1}\right)$, so that
the subleading correction becomes smallest for $S_{N-1}=0$ and 
$S_{N-2}\approx\frac{N}{2}d^{N-2}$. Substituting this result into 
Eq.~\eqref{eq:2-purity} leads to $S_2\approx \frac{N}{2}d^2$. In particular
the latter requirement together with $S_1=0$ can be fulfilled 
if the state is a {\sf Bell} tensor product. 

For the opposite limit, $N\gg d > 2$, 
general statements are more difficult to make, because
the sector sum does not correspond to a power expansion in $d$ any longer.
We can discuss at least the case 
of states that are more entangled than GHZ, that is, 
$m$-uniform states~\cite{Scott2004,Goyeneche2014}. A state is called
$m$-uniform if $S_1=S_2=\cdots =S_m=0$, with the extreme case of 
AME states ($m=\lfloor N/2\rfloor$).
For AME states, $S_N\simeq d^N(1-\frac{1}{d^2})^N$ is a fair approximation
that applies to some extent also to 
$m\lesssim N/2$ if $N$ does not exceed $d^2$. 
Then, for large $d$  approximately $S_N \sim d^N \rme^{-\frac{N}{d^2}}$, which
shows that a substantial fraction of the Bloch vector length is not in the
$N$-sector, making these states bad candidates for the maximum $S_N$.

On the other hand, for the GHZ state we have 
\begin{align}
S_N(\text{GHZ}_d^N)\simeq d^N\left(1-\frac{1}{d}\right) + 
                                  d^{N-1}\rme^{-\frac{N}{d}}
\ \ .
\label{eq:ghzlimit}
\end{align}
That is, $S_N$ is essentially given by the first term in Eq.~\eqref{eq:ghzlimit}
and the relative error shrinks with increasing $N$. 
This is the expected
behavior for the dominating state, because $d^N\left(1-\frac{1}{d}\right)$
is the absolute maximum the traceless part of a pure state 
$\ket{\psi}\!\bra{\psi}$ can achieve:
An offdiagonal element consisting of orthogonal product states gives
$\| \ket{jj\ldots j}\!\bra{kk\ldots k} \|_{\mathcal{P}} = 1$, which
is the maximum  among all rank-1 operators. 
An $N$-qudit state
of local dimension $d$ can have at most Schmidt
rank $d$ in a bipartition of a single party against the rest.
This amounts 
to a maximum offdiagonal contribution of $ \frac{1}{d^2}d(d-1) d^N$ to $S_N$ --
which is precisely the GHZ result. Consequently, for large $N$ the GHZ
state $N$-sector approaches the maximum for any rank-1 operator.
Evidently, this discussion cannot exclude the existence of a
state that approaches this maximum even faster.
%
%

\section{Conclusions}
%
We have analyzed the Bloch sector 
distribution for multipartite pure quantum states
of $N$ $d$-level systems, in particular the properties of the $N$-sector.
We have demonstrated that the sectors must not be considered individually;
rather, there are numerous interdependencies that determine the
distribution. One of our main results based on this insight
is the proof that for qubits
the GHZ state has maximum $N$-sector also for even $N$.
We have given an extensive characterization of the $N$-sector behavior
for arbitrary $N$ and $d$, which can be viewed as an algebraic
problem analogous to that of the existence of AME states.
Most importantly, we find that strong $N$-party correlations (viz
maximum $N$-sector) do {\em not} necessarily imply genuine 
multipartite entanglement. Apart from our physics results, 
our work provides several novel technical tools for 
analyzing the Bloch representation of pure
states and thereby shows that this is a powerful approach
to obtain new insight into the 
mathematical properties of many-body quantum states.

\acknowledgments
This work was funded by the German Research Foundation Project
EL710/2-1 (C.E., J.S.), by Grant PGC2018-101355-B-100 (MCIU/FEDER/UE) and
Basque Government Grant IT986-16 (J.S.).
The authors would like to thank Marcus Huber and Nikolai Wyderka
for stimulating discussions. C.E.\ and J.S.\ acknowledge
Klaus Richter's support of this project. 
%
%

%

\section*{Appendix}
\renewcommand{\thesection}{\Alph{section}}
\renewcommand{\theequation}{\thesection\arabic{equation}}
\setcounter{equation}{0}
\setcounter{section}{1}
%
\subsection*{Sector distribution of GHZ state, Eq.~(8)}
%
In order to obtain the sector lengths for the GHZ state it is not
necessary to explicitly calculate the Bloch representation.
Yet we quickly do it for the qubit example  to demonstrate how simple
it is. The density matrix of the $N$-qubit GHZ state is
\begin{align}
 \Pi_{\mathrm{GHZ}_2^N} &\ =\  \frac{1}{2}\bigg(
                        \ket{00\ldots 0}\!\bra{00\ldots 0}\ + \bigg.
\nonumber\\
                        &\ \ \ \ \ \ \ \ \ \bigg. +
                        \ket{11\ldots 1}\!\bra{11\ldots 1}\ +
\nonumber\\
              &  \ \ \ \ \ \ \ \ \ \ \      +\ket{00\ldots 0}\!\bra{11\ldots 1}
                 \ +
\nonumber\\
              &\ \ \ \ \ \ \ \ \ \ \ \ \   +   \bigg.      
               \ket{11\ldots 1}\!\bra{00\ldots 0}
                                   \bigg) \ .
\end{align}
Each term here is a tensor product of $N$ identical rank-1
single-qubit operators, 
\begin{align*}
 \ket{0}\!\bra{0}^{\otimes N}\ =\ &\frac{1}{2^N}(\id+Z)^{\otimes N}
\\
 \ket{1}\!\bra{1}^{\otimes N}\ =\ &\frac{1}{2^N}(\id-Z)^{\otimes N}
\\
 \ket{0}\!\bra{1}^{\otimes N}\ =\ &\frac{1}{2^N}
                                      (X+\rmi Y)^{\otimes N}
\\
 \ket{1}\!\bra{0}^{\otimes N}\ =\ &\frac{1}{2^N}
                                      (X-\rmi Y)^{\otimes N}
\ \ ,
\end{align*}
where $X\equiv\sigma_1$, $Y\equiv\sigma_2$, $Z\equiv\sigma_3$ 
are the Pauli matrices and $\id$ is the qubit identity matrix. Hence
\begin{align}
 & \Pi_{\mathrm{GHZ}_2^N} \ = \  \frac{1}{2^N}\bigg(
                \sum_{\mathrm{even} \# Z} ZZ\ldots \id\id\ +
\nonumber\\
 &  \ \ \ \ \ \    + \sum_{\mathrm{even \# Y}} (-1)^{\frac{\# Y}{2}}
                           X\ldots YY\ldots X\ldots YY
                                   \bigg) \ \ ,
\end{align}
where the sums run over all combinations of even numbers of $Z$ occurrences
padded with $\id$s (diagonal) and even numbers of $Y$ occurences padded
with $X$s (offdiagonal); for simplicity we omit the tensor product signs.
The difference between even and odd $N$ is that for even $N$ there is 
one $N$-sector term $ZZ\ldots Z$ in the diagonal part, whereas for odd $N$
the $N$-sector exclusively consists of offdiagonal terms.
The GHZ state for $d>2$ can be built in an analogous manner.

In order to derive Eq.~\eqref{eq:ghz-sectors} we can take a shortcut
and use Eq.~\eqref{eq:NSecPurity},
\begin{align}
   S_N = d^N\sum_{A} \left(-\frac{1}{d}\right)^{|A|} 
                            \Tr\left(\rho_{\bar{A}}^2\right)\ .
\end{align}
The GHZ state is particularly simple as all reduced states are
of rank $d$ and completely mixed on their span, so that
$\Tr\left(\rho_A^2\right)=\frac{1}{d}$ for all $|A|\neq 0, N$.
For $|A|=0$ and $|A|=N$ we have 
$\Tr\left(\rho_A\right)^2=1$, so that
\begin{align}
    S_N  =  & d^{N}\frac{1}{d}\left(1-\frac{1}{d}\right)^N
             +d^N\left(1-\frac{1}{d}\right) -
\nonumber\\
          & -(-1)^N d^N\left(\frac{1}{d^{N+1}}-\frac{1}{d^{N}}\right) 
\nonumber\\
          =  & d^{N-1}(d-1)\  +\ 
\nonumber\\
             & \ \  +\ \frac{1}{d}\left[(d-1)^N
           +(-1)^N (d-1)\right] \ .
\end{align}
For $k<N$,  $S_k$ is given by the length of the 
last sector of the reduced density matrix $\rho_A$ ($|A|=k$)
times the number of such reduced density matrices, 
$S_k=d^k {N\choose k}\|\rho_A\|_{\mathcal{P}}^2$. In contrast to
the $N$-sector we need not include a correction for the first
term, so that
%
\begin{align}
    S_k\  = & \  d^{k}\frac{1}{d}\left(1-\frac{1}{d}\right)^k{N\choose k} -
\nonumber\\ & \ \ \ \ \ \ \ \ \ \ \ \ \ \ 
           -(-1)^k d^k\left(\frac{1}{d^k}-\frac{1}{d^{k-1}}\right) 
\nonumber\\
          = &   {N\choose k}\frac{1}{d}\left[(d-1)^k
           +(-1)^k (d-1)\right]  \ .
\end{align}
%

\subsection*{Proof maximum $N$ sector of odd-$N$ qubit GHZ state}
Here we show the proof that for odd $N$ qubits, the maximum $N$-sector length
is  $\max S_N=S_N(\mathrm{GHZ})=2^{N-1}$, which is realized 
by the GHZ state~\cite{Tran2016}.

First, we recall that for odd $N$ qubit states $\Pi=\ket{\psi}\!\bra{\psi}$
the degree-2 SL invariant
\begin{align}
H\ =\ \Tr\left[ \Pi\ Y^{\otimes N}\Pi^*Y^{\otimes N}\right]
\ =\ 0
\label{eq:H}
\end{align}
always vanishes~\cite{ES2015,Tran2016,Wyderka2018} (here, $Y\equiv\sigma_2$ is 
a Pauli matrix and $\Pi^*=\ket{\psi^*}\!\bra{\psi^*}$, 
where $\ket{\psi^*}$ is the vector
with complex conjugate components). In terms of sector lengths
Eq.~\eqref{eq:H} reads~\cite{ES2015,Tran2016,Wyderka2018} 
$0=\sum_{k=0}^N (-1)^k S_k$, so that
\begin{align}
    S_{\mathrm{even}}\ \equiv\ \sum_{k=0}^{\frac{N-1}{2}} S_{2k}\ =\ 
    \sum_{k=0}^{\frac{N-1}{2}} S_{2k+1}\ \equiv\ S_{\mathrm{odd}}\ \ ,
\label{eq:even=odd}
\end{align}
that is, the sum of the even-numbered sector lengths $S_{\mathrm{even}}$
{\em always equals}
that of the odd-numbered ones, $S_{\mathrm{odd}}$. 
Because of the purity-0 constraint
$\sum_{k=0}^N S_k =S_{\mathrm{even}}+S_{\mathrm{odd}}= 2^N$ 
this means that both even and odd sector length
sums are always equal to $2^{N-1}$, so that the properties of a state
are encoded in the distribution of the even sector lengths 
among themselves on the one hand, and separately the odd ones, 
on the other hand.

It is quite obvious then that the GHZ state (odd $N$) is the one with
maximum $N$-sector: Here, the entire odd sector length $2^{N-1}$ is shifted
to the $N$-sector, and the other odd sector lengths vanish. (The peculiarity
is that such a state actually does exist -- this is by no means
guaranteed by  Eq.~\eqref{eq:even=odd} and the purity
constraint.)
Note also that for the GHZ state, 
Eq.~\eqref{eq:even=odd} does not say anything about
the distribution of the even-numbered sectors.

Now consider the $\mathcal{PQ}$ relation, Eq.~\eqref{eq:PQrel} from the
main text,
\begin{align}
 S_{N} & \ =\   \frac{1}{2^N} \sum_{k=0}^{N}(-1)^k 3^{N-k} S_{k}
\nonumber\\
            \  =\  &\frac{1}{2^N} \bigg[
     \sum_{l=0}^{\frac{N-1}{2}} 3^{N-2l}S_{2l}\ -\ 
          \sum_{m=0}^{\frac{N-1}{2}} 3^{N-2m-1}S_{2m+1}
                                 \bigg]
\  .
\label{eq:PQrel2}
\end{align}
We see that also from the point of view of Eq.~\eqref{eq:PQrel2}
the maximum $N$-sector for the GHZ state makes perfect sense: 
All odd sector contributions are moved to $S_{N}$ where they cause
the `least damage' for maximizing the r.h.s.\ of the equation, because
$S_N$ has the smallest prefactor.

\subsection*{Proof maximum l.h.s.\ of Eq.~\eqref{eq:evenlin} in main text}
In the following we demonstrate the last step of the proof
in the main text that $\max S_N=2^{N-1}+1$ for even-$N$ qubit GHZ states.
This step consists in maximizing the left-hand side (l.h.s.) of
Eq.~\eqref{eq:evenlin} of the manuscript,
\[
       S_1 +  \sum_j 2 \Tr R_{[j]}\ \longrightarrow\ \max\ \ ,
\]
where $\rho_{[j]}=\Tr_{\{j\}}\Pi_x$ and $\Pi_x=\ket{x}\!\bra{x}$ 
is a pure state.

First, let us consider $\rho_{[1]}=\Tr_{\{1\}}\Pi_x$.
We write the Schmidt decomposition of $\ket{x}$ with respect to the
first qubit,
\begin{align}
    \ket{x}\ =\ \sqrt{\lambda_1}\ket{0}\ket{X_0}+\sqrt{1-\lambda_1}
                                \ket{1}\ket{X_1}\ \ ,
\end{align}
where $\{\ket{0},\ket{1}\}$ is the Schmidt basis on the first qubit
and $\ket{X_0}$, $\ket{X_1}$ two orthogonal odd-$(N-1)$ qubit states,
the Schmidt vectors on qubits $\{2 \ldots N\}$. Hence,
\begin{align}
      \rho_{[1]}\ =\ \lambda_1\ket{X_0}\!\bra{X_0}+
                     (1-\lambda_1)\ket{X_1}\!\bra{X_1}
\ \ ,
\end{align}
so that, as for $k$-qubit states $\ket{\phi}$ the inverted
state \mbox{$|\tilde{\phi}\rangle=Y^{\otimes k}\ket{\phi^*}$}
[cf.~\cite{ES2015,ES2018,Wyderka2018} and the discussion below 
Eq.~\eqref{eq:H}],
\begin{align}
      \Tr R_{[1]}\ =\ & \Tr\left[\rho_{[1]}\ Y^{\otimes(N-1)}
                               \rho_{[1]}^* Y^{\otimes(N-1)}\right]
\nonumber\\
                   =\ & \lambda_1^2| \langle X_0|\tilde{X}_0\rangle|^2+
                        (1-\lambda_1)^2\langle X_1|\tilde{X}_1\rangle|^2+
\nonumber\\           & +\lambda_1(1-\lambda_1) 
                         |\langle X_0 | \tilde{X}_1\rangle|^2+
\nonumber\\           & +\lambda_1(1-\lambda_1) 
                         |\langle X_1 | \tilde{X}_0\rangle|^2
\nonumber\\     
                   =\ & 2\lambda_1(1-\lambda_1) 
                         |\langle X_0 | \tilde{X}_1\rangle|^2
\nonumber\\
                   =\ & 2\lambda_1(1-\lambda_1) 
                         |\langle X_0 | Y^{\otimes (N-1)} X_1^*\rangle|^2
\ \ ,
\label{eq:TrR1}
\end{align}
because $\langle \phi | \tilde{\phi}\rangle = 0$ for odd-$N$ qubit states
[see also Eq.~\eqref{eq:H}]. For the matrix element in Eq.~\eqref{eq:TrR1}
we have
\begin{align}
   \Delta\ =\ |\langle X_0 | Y^{\otimes (N-1)} X_1^*\rangle|\ \leqq\ 1\ \ ,
\end{align}
since the operator $Y^{\otimes (N-1)}$ has only eigenvalues of modulus 1.
Consequently we find, if we add the $1$-sector $S_1^{(1)}$ of the 1st qubit
in $\Pi_x$,
\begin{align}
   & \max_{\ket{x}}  \left[ S_1^{(1)} + 2\Tr R_{[1]}  \right] \  = \ 
\nonumber\\
           & =   \max_{\lambda_1,\Delta}\ \left[
                                    2\lambda_1^2+2(1-\lambda_1)^2-1
                                   + 4\lambda_1(1-\lambda_1) \Delta^2
                                \right]
\nonumber\\    & =  1\ \ .
\label{eq:max1}
\end{align}
The states $\ket{X_0}$, $\ket{X_1}$ that realize this maximum are, e.g.,
$\ket{X_0}=\ket{0}^{\otimes (N-1)}$ and
$\ket{X_1}=\ket{1}^{\otimes (N-1)}$. That is, an even-$N$ qubit state
that maximizes the l.h.s.\ of Eq.~\eqref{eq:max1} is, e.g., the GHZ state
$\ket{x}=\frac{1}{\sqrt{2}}\left(\ket{0}^{\otimes N}+
                                 \ket{1}^{\otimes N}\right)$.

The same reasoning as above can be applied for all qubits $j=1\ldots N$,
so that we find for the symmetrized l.h.s.\ of Eq.~\eqref{eq:max1},
\begin{align}
         \max_{\ket{x}}\left[
           S_1 + \sum_j 2 \Tr R_{[j]}\right]\ =\ N\ \ ,
\end{align}
with the GHZ state attaining the maximum.

%


\begin{thebibliography}{99}
\bibitem{footnote2}
An in-depth analysis of this projection operator
will be carried out in forthcoming work.
\end{thebibliography}
\end{document}